\documentclass[journal=jacsat,manuscript=article]{achemso}

\usepackage[version=3]{mhchem} 
\usepackage{float} 
\usepackage{amssymb} 

\author{Shagun Maheshwari}
\affiliation{Department of Materials Science Engineering, Carnegie Mellon University, 5000 Forbes Avenue, Pittsburgh, PA 15213, USA}
\altaffiliation{These authors contributed equally to this work.}
\author{Zhengxian Tang}
\affiliation{Department of Materials Science Engineering, Carnegie Mellon University, 5000 Forbes Avenue, Pittsburgh, PA 15213, USA}
\altaffiliation{These authors contributed equally to this work.}
\author{Janghoon Ock}
\affiliation{Department of Chemical Engineering, Carnegie Mellon University, 5000 Forbes Avenue, Pittsburgh, PA 15213, USA}
\alsoaffiliation[unl-cheme]
{Department of Chemical and Biomolecular Engineering, University of Nebraska--Lincoln, Lincoln, NE 68588, USA}
\author{Adeesh Kolluru}
\affiliation{Department of Chemical Engineering, Carnegie Mellon University, 5000 Forbes Avenue, Pittsburgh, PA 15213, USA}
\author{John R. Kitchin}
\affiliation{Department of Chemical Engineering, Carnegie Mellon University, 5000 Forbes Avenue, Pittsburgh, PA 15213, USA}
\author{Amir Barati Farimani}
\affiliation{Department of Mechanical Engineering, Carnegie Mellon University, 5000 Forbes Avenue, Pittsburgh, PA 15213, USA}
\email{barati@cmu.edu}

\title[An \textsf{achemso} demo]
  {Beyond Force Metrics: Pre-Training MLFFs for Stable MD Simulations}

\abbreviations{IR,NMR,UV}

\begin{document}



\begin{abstract}
Machine-learning force fields (MLFFs) have emerged as a promising solution for speeding up \emph{ab initio} molecular dynamics (MD) simulations, where accurate force predictions are critical but often computationally expensive. In this work, we employ GemNet-T, a graph neural network model, as an MLFF and investigate two training strategies: (1) direct training on MD17 (10K samples) without pre-training, and (2) pre-training on the large-scale OC20 dataset followed by fine-tuning on MD17 (10K). While both approaches achieve low force mean absolute errors (MAEs), reaching 5 meV/Å per atom, we find that lower force errors do not necessarily guarantee stable MD simulations. Notably, the pre-trained GemNet-T model yields significantly improved simulation stability, sustaining trajectories up to three times longer than the model trained from scratch. By analyzing local properties of the learned force fields, we find that pre-training produces more structured latent representations, smoother force responses to local geometric changes, and more consistent force differences between nearby configurations, all of which contribute to more stable and reliable MD simulations. These findings underscore the value of pre-training on large, diverse datasets to capture complex molecular interactions and highlight that force MAE alone is not always a sufficient metric of MD simulation stability.
\end{abstract}

\section{Introduction}
Molecular dynamics (MD) simulation is a computational method used to analyze the physical movements of atoms and molecules over time, enabling the exploration of molecular interactions, structural rearrangements, and thermodynamic properties at the atomic scale. The primary aim of MD simulations is to provide insights into atomistic processes that are challenging or impossible to observe experimentally. Typical examples include folding pathways of small proteins and peptides \cite{Lindorff-Larsen2011}, ligand binding kinetics in drug discovery campaigns \cite{nunesalves2020binding}, ion transport in solid-state battery electrolytes \cite{mabuchi2021ion}, crack propagation in structural alloys \cite{lee2023crack}, and nucleation events in super-cooled liquids \cite{sosso2016nucleation}.

Density functional theory (DFT) plays a critical role in MD simulations by providing quantum mechanical calculations of interatomic forces derived from the electronic structure. While DFT yields highly accurate forces by explicitly accounting for electron interactions, its computational cost scales poorly with system size and simulation time, limiting its applicability to large-scale or long-timescale simulations \cite{unke2021mlff}. This limitation renders DFT impractical for long MD trajectories involving hundreds of thousands of time-steps or for simulations of systems with thousands of atoms. For this reason, classical MD simulations often rely on empirical force fields that approximate these interactions, sacrificing some accuracy to achieve feasible simulation durations and system sizes
\cite{smith2018ani1, gunceler2023improving}.

To address the computational limitations of \emph{ab initio} methods like DFT, MLFFs have emerged as promising alternatives for MD simulations \cite{unke2021mlff}. MLFFs are data-driven models trained to approximate the potential energy surface (PES) and interatomic forces of atomistic systems based on reference calculations from quantum mechanical methods like DFT. By learning the mapping from atomic configurations to energies and forces, MLFFs enable simulations that approach quantum-level accuracy while being orders of magnitude faster than traditional \emph{ab initio} approaches. As a result, they allow the exploration of larger systems and longer timescales that would otherwise be intractable using direct quantum mechanical methods.

Significant advancements have been made in the field of MLFFs, primarily through graph neural network (GNN) models such as SchNet \cite{schutt2017schnet}, DimeNet \cite{klicpera2020dimenet}, PaiNN \cite{schutt2021painn}, DeepPot-SE \cite{zhang2018deeppot}, NequIP \cite{batzner2022nequip}, and GemNet \cite{gasteiger2021gemnet}. These models leverage graph-based representations where atoms are treated as nodes and atomic interactions as edges. SchNet introduced continuous convolutional filters to model atomic interactions, while DimeNet extended these ideas by explicitly incorporating angular information. PaiNN further refined the approach by utilizing equivariant message-passing layers that respect rotational symmetries. GemNet advanced this by incorporating directional embeddings to better capture radial and angular geometric features, providing superior handling of directional information crucial for accurate molecular force prediction.

Despite their successes, MLFFs exhibit notable shortcomings, particularly regarding simulation stability and extrapolation behavior. A benchmark study conducted by Fu et al.\cite{fu2023forces} emphasizes the limitations of force prediction metrics alone, demonstrating that high accuracy in force prediction does not always correlate with stable and realistic MD trajectories. They observed that models frequently failed simulations due to instability caused by extrapolation errors, where models predicted extreme, non-physical forces for configurations inadequately represented in training datasets. This instability often resulted in catastrophic trajectory divergence, bond breakage, and unrealistic configurations, even if models initially exhibited low force prediction errors. Fu et al.\cite{fu2023forces} highlighted that these failures predominantly occur in simulations requiring long timescales or involving systems with complex interactions not well captured by the training data. These findings highlight stability and comprehensive configurational space coverage as critical metrics, suggesting that the development of robust MLFFs requires training on diverse molecular configurations to improve both simulation stability and accuracy.


Motivated by these limitations, we explored pre-training strategies for improving MLFF performance and stability, with the goal of learning better molecular representations and more accurate correlations among similar structures. Prior studies have shown substantial benefits of pre-training, where initial training on large and diverse datasets enhances the transferability and robustness of MLFFs. For example, pre-training has been shown to significantly enhance the generalizability and stability of models, allowing them to better capture complex molecular interactions encountered in extended simulations \cite{chanussot2021oc20, raja2025stabilityawaretrainingmachinelearning, wang2025pfdautomaticallygeneratingmachine, peptidebert}. Additionally, recent work has shown that pretrained models, such as EquiformerV2, can be fine-tuned to achieve state-of-the-art performance even on challenging, out-of-domain systems like high-entropy alloys, significantly accelerating atomistic discovery in materials science \cite{clausen2024}. More broadly, various forms of pre-training allow graph neural networks to develop transferable atomic representations across diverse domains. These include small molecules, catalysts, and inorganic crystals, fostering more universal chemical understanding \cite{Wang2022, Magar2022, ock2024multimodal}.

In this study, we investigate how pre-training an MLFF, specifically GemNet-T, on large-scale datasets impacts the stability of MD simulations. We show that pre-training on diverse atomic systems, including catalyst datasets, followed by fine-tuning on smaller molecular datasets, effectively reduces overfitting and enhances generalizability. Pre-training yields more structured latent representations, smoother force responses to local geometric perturbations, and more consistent force differences between nearby configurations. As a result, model stability is substantially improved, even though the force prediction accuracy itself does not change significantly. These results underscore the significance of comprehensive, large-scale pre-training in capturing complex molecular interactions and provide a foundational framework for developing robust MLFFs for MD simulations.

\section{Methods}

\subsection{Molecular Dynamics Simulation}

MD simulation is a computational method used to model the physical movements of atoms and molecules over time by solving Newton’s equations of motion\cite{frenkel2001understanding}. The MD simulation follows the Newtonian equation of motion, given by:

\begin{equation}
    \frac{d^2 \mathbf{x}}{dt^2} = m^{-1} \mathbf{F}(\mathbf{x}),
\end{equation}

where \( \mathbf{x} \) denotes the atomic positions, \( m \) is the mass of the atom, and \( \mathbf{F}(\mathbf{x}) \) represents the force derived from the learned potential energy. The forces are calculated as the negative gradient of the predicted potential energy:

\begin{equation}
    \mathbf{F}(\mathbf{x}) = - \nabla E(\mathbf{x}).
\end{equation}

The atomic positions at each time step \( t \) are updated using the velocity Verlet integration scheme:

\begin{equation}
    \mathbf{r}_i(t + \Delta t) = \mathbf{r}_i(t) + \mathbf{v}_i(t) \Delta t + \frac{1}{2} \mathbf{a}_i(t) (\Delta t)^2,
\end{equation}

where \( \mathbf{r}_i(t) \) is the position of atom \( i \) at time \( t \), \( \mathbf{v}_i(t) \) is the velocity, \( \mathbf{a}_i(t) \) is the acceleration derived from the predicted forces, and \( \Delta t \) is the time step.

To control the temperature, a Nos\'e--Hoover thermostat was employed to maintain a target temperature of 500~K \cite{fu2023forces}. The simulation was conducted using a time step of 0.5~fs and was run for 600,000 steps, corresponding to 300~ps of dynamics. Atomic positions were periodically saved for subsequent analysis. 

By combining the Nos\'e--Hoover thermostat with the velocity Verlet integrator, the MD simulation effectively captured thermal fluctuations while preserving the physical consistency imposed by the machine-learned potential. This procedure generated a time series of atomic positions and velocities, \( \{\mathbf{x}_t\}_{t=0}^{T} \), enabling a detailed evaluation of the system's dynamical behavior under the trained model.

\subsection{Structural Validation}

To assess the structural fidelity of MLFFs in MD simulations of small molecules like Aspirin, we compute the pair-distance distribution function, \( h(r) \). Unlike the radial distribution function (RDF) used in bulk or periodic systems, the pair-distance distribution function is better suited for finite molecular systems \cite{fu2023forces}, capturing the distribution of all interatomic distances within a molecule. It is formally defined as:

\[
h(r) = \frac{1}{N(N-1)} \sum_{i=1}^{N} \sum_{j \ne i} \delta(r - \| \mathbf{x}_i - \mathbf{x}_j \|),
\]

where \( N \) is the number of atoms and \( \mathbf{x}_i \) are their positions. This function provides a low-dimensional, geometry-sensitive descriptor of molecular structure, enabling direct comparisons between model-predicted and reference (\textit{ab initio}) simulations. Following prior work~\cite{raja2025stabilityawaretrainingmachinelearning}, we are also able to compute the MAE between predicted and reference pair-distance distribution function curves to quantify structural deviations:

\[
\text{MAE}_{h(r)} = \int_0^\infty \left| \langle h(r) \rangle - \langle \hat{h}(r) \rangle \right| dr.
\]

This metric helps detect unphysical distortions like bond stretching or angle collapse, common causes of failure in MLFF-driven MD, providing a more interpretable and structure-specific complement to conventional force MAE.

\subsection{Stability Criterion}

We assess the stability of MD simulations for flexible molecules by monitoring bond lengths over time. A simulation is classified as unstable if, at any time \( t < 300\,\text{ps} \), the bond length between any atom pair exceeds a specified threshold, indicating bond breaking or excessive stretching that compromises structural integrity before the intended simulation duration. The criterion for instability is expressed as \cite{fu2023forces}:

\begin{equation}
\max_{(i,j) \in \mathcal{B}} 
\left| \|\mathbf{x}_i(t) - \mathbf{x}_j(t)\| - b_{i,j} \right| > \Delta
\end{equation}

The time at which the simulation first exhibits a loss of structural integrity is defined as the instability onset time, denoted \( t_{\mathrm{inst}} \), and is given by:

\begin{equation}
t_{\mathrm{inst}} = \min \left\{ t \in [0, T] : \max_{(i,j) \in \mathcal{B}} 
\left| \|\mathbf{x}_i(t) - \mathbf{x}_j(t)\| - b_{i,j} \right| > \Delta \right\}
\end{equation}

where \( \mathcal{B} \) is the set of bonded atom pairs, \( \|\mathbf{x}_i(t) - \mathbf{x}_j(t)\| \) denotes the distance between atoms \( i \) and \( j \) at time \( t \), \( b_{i,j} \) is the reference bond length for the atom pair, and \( \Delta \) is the allowable deviation threshold, set to 0.5 Å.

\subsection{Machine Learning Force Fields}

MLFFs learn an approximate representation of the potential energy surface from reference data, typically generated by quantum mechanical methods such as DFT. At each simulation step, the current atomic positions are input to the MLFF, which predicts the forces acting on each atom. These predicted forces are then used to iteratively update atomic positions throughout the simulation. Recently, GNN architectures have gained prominence for effectively capturing atomic connectivity in molecular systems. GNNs capture the connectivity of atomic structures by representing 3D atomic systems as graphs, where atoms are treated as nodes and interatomic distances define the edges \cite{gasteiger2021gemnet}.

In this study, we utilize GemNet-T as the MLFF model. GemNet-T is a high-performing GNN designed to learn molecular energies and forces, building on prior models like DimeNet++ \cite{gasteiger2021gemnet}. It enhances the representation of multi-scale geometric interactions by incorporating directional embeddings, capturing both radial and angular features. This enables more accurate force predictions, particularly for complex three-dimensional atomic arrangements. 

Figure~\ref{fig:framework}c illustrates the core structure of the GemNet-T model, a directional message-passing neural network designed for MD simulations. The model begins by embedding atomic information using both atomic numbers and local geometric features, including distances, angles, and dihedral angles. These inputs generate initial atom and edge embeddings that preserve the molecular structure. The embeddings are then processed through multiple stacked interaction blocks, each consisting of a directional message-passing module and an atom-wise update. In GemNet-T, the message passing operates over one-hop neighborhoods and uses triplet-based geometric information (from distance and angle inputs) to encode directional relationships. This approach avoids the complexity of higher-order interactions, such as the quadruplet-based dihedral terms used in GemNet-Q, while still capturing essential directional features~\cite{gasteiger2021gemnet}. After each message-passing step, atomic embeddings are updated through residual connections. Once all interaction layers are completed, the final atomic embeddings are aggregated to predict molecular energies and atomic forces. Importantly, these predictions are designed to be equivariant to translations, rotations, and reflections, ensuring that the model produces physically consistent results during MD simulations~\cite{gasteiger2021gemnet}.

\subsection{Datasets}
We use the MD17 dataset as the primary benchmark for small-molecule systems. MD17 consists of \emph{ab initio} MD trajectories, where each snapshot provides atomic coordinates, DFT total energy, and per-atom forces \cite{chmiela2017machine}. Aspirin, a medium-sized organic molecule with 21 atoms and multiple rotatable bonds, is selected as a representative system due to its structural flexibility. This flexibility amplifies the impact of small force prediction errors, potentially leading to unstable trajectories over time. The aspirin subset in MD17 includes over 20,000 configurations (i.e., molecular snapshots along an MD trajectory, each with atomic positions, energy, and forces). For this study, we use 9,500 configurations for training, 500 for validation, and 10,000 as a separate test set. 

To extend the model's generalization beyond isolated small molecules, we incorporate the Open Catalyst 2020 (OC20) dataset for pre-training. OC20 contains 1.3 million relaxations of molecular adsorptions on surfaces, supported by over 260 million corresponding DFT calculations \cite{chanussot2021open}. We specifically utilize a 2-million-sample subset from the Structure-to-Energy-and-Force (S2EF) task, which provides forces and energies for each relaxed configuration. Unlike MD17, OC20 data include periodic boundary conditions, more accurately reflecting realistic catalytic surfaces. By training on this larger, chemically diverse dataset, we aim to equip the model with robust atomic interaction representations before fine-tuning on the smaller MD17 aspirin set.

\section{Experiments}

\begin{figure}[!htb]
    \centering
    \includegraphics[width=0.99\textwidth]{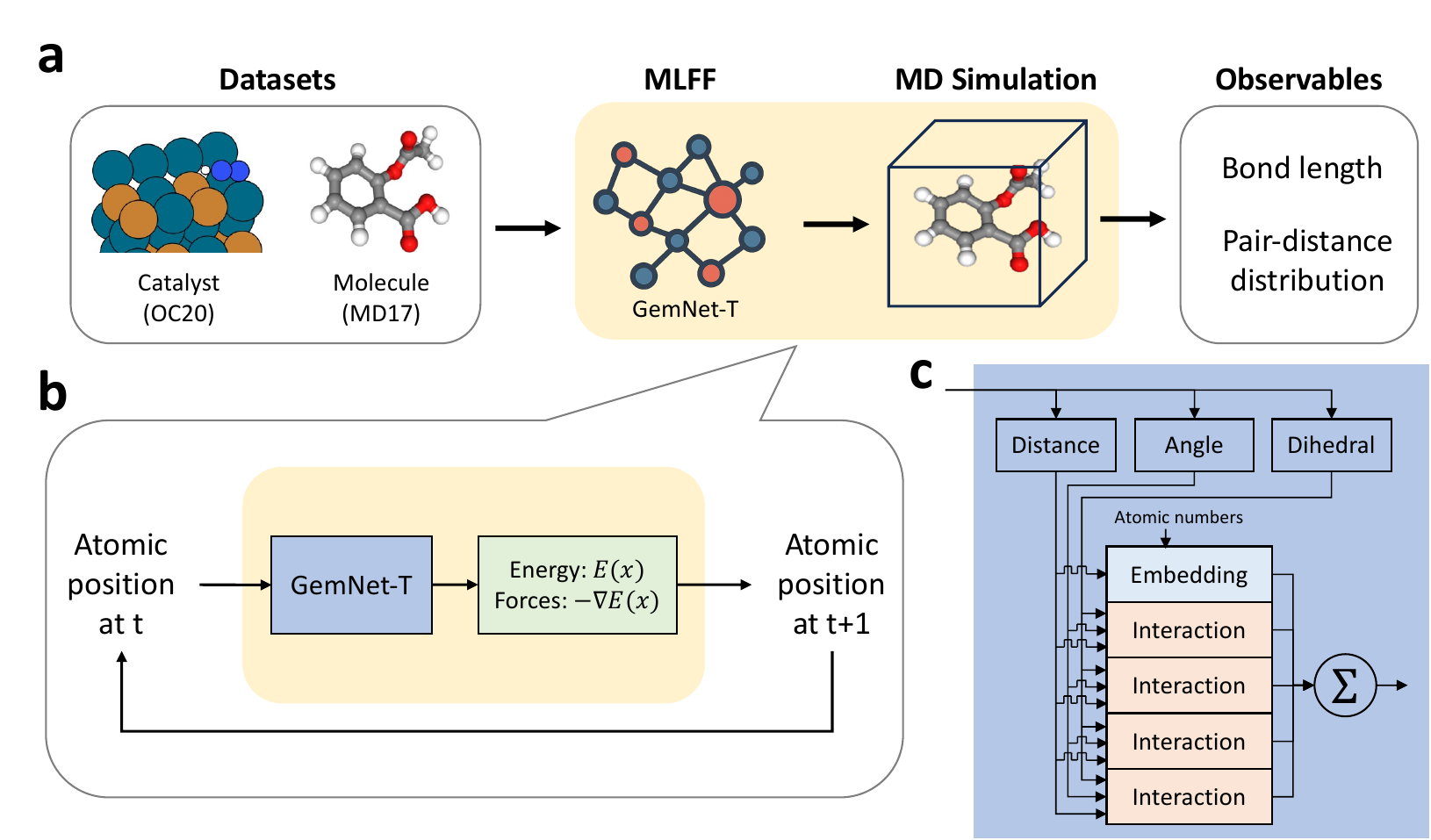}
    \caption{Overview of the MLFF framework for MD simulations.  
    \textbf{a.} GemNet-T is trained on molecular (MD17) and catalyst (OC20) datasets, then used as a force predictor in MD simulations.  
    \textbf{b.} Atomic positions are iteratively updated based on forces predicted by the trained GemNet-T model.  
    \textbf{c.} Simplified illustration of the GemNet-T architecture, which incorporates multi-hop geometric message passing and directional embeddings to model atomic interactions.}
    \label{fig:framework}
\end{figure}

The trained GemNet-T model serves as the MLFF for MD simulations of the aspirin molecule. Simulation stability is assessed by analyzing bond length divergence using \emph{ab initio} MD data from the MD17 dataset, as depicted in Figure~\ref{fig:framework}a. During the simulation, the MLFF takes the structure from each simulation step and predicts the atomic forces to update atomic positions for the next step, as illustrated in Figure~\ref{fig:framework}b. Simplified illustration of the GemNet-T model is provided in Figure~\ref{fig:framework}c.

To enhance MD stability, we adopt a two-stage training strategy. Specifically, we fine-tune a GemNet-T model that was initially pre-trained on the large and chemically diverse OC20 dataset, using the target molecule aspirin from the MD17 dataset. We compare both the force prediction accuracy and the resulting MD simulation stability of this pre-trained model against a baseline model trained from scratch using only the aspirin dataset.

For the aspirin-only training, the GemNet-T model is configured with four message-passing layers, a 128-dimensional atom/edge embedding size, and is optimized using AdamW. In the pre-training strategy, we utilize a GemNet-T model pre-trained on the \texttt{OC20-S2EF-2M} subset, a 2-million structure dataset derived from 1.3 million relaxation trajectories of adsorbate–catalyst systems in the OC20 benchmark.
The pre-trained model employs a 512-dimensional atom/edge embedding size and is subsequently fine-tuned on the aspirin subset from MD17.  


\section{Results and Discussion}
\subsection{Force Metric Limitations}

\begin{figure}[!htb]
    \centering
    \includegraphics[width=0.99\textwidth]{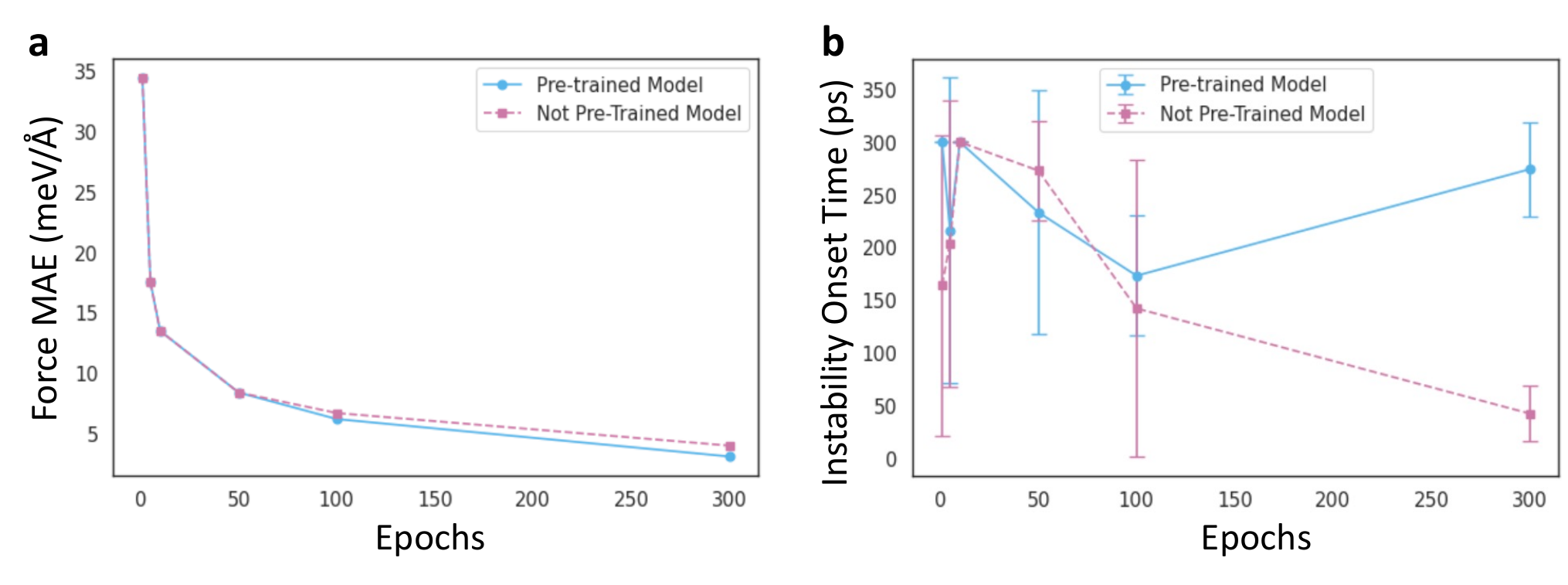}
    \caption{Impact of pre-training on force accuracy and simulation stability.  
\textbf{a.} Force MAE (meV/\AA) over epochs for the GemNet-T model trained without pre-training (MD17 only) versus with pre-training (MD17 pre-training followed by OC20 finetuning).  
\textbf{b.} Instability onset time over epochs. Error bars represent results from three independent simulations.}
    \label{fig:results}
\end{figure}

We evaluate the stability of the MD simulations to determine whether high force prediction accuracy necessarily guarantees more stable simulation results. The MAE in force prediction serves as the primary error metric during MLFF training, as it is the objective function the model is optimized for. To evaluate dynamical stability, we use the instability onset time (\( t_{\mathrm{inst}} \)), which measures how long the simulation preserves structural integrity before any bond deviates beyond a defined threshold from its reference length. A higher instability onset time, approaching the 300~ps reference from \emph{ab initio} MD, indicates greater stability and closer agreement with the ground truth trajectory. The instability onset times shown in Figure~\ref{fig:results}b represent the average values obtained from three independent simulations initialized with different random seeds.

As shown in Figure~\ref{fig:results}a, the force MAE curve decreases smoothly as training progresses. However, the difference between the pre-trained and non-pre-trained cases is minimal with a $\text{RMS Deviation} \approx 0.421\ \text{meV/\AA}$ at Epoch 300. This result makes it difficult to distinguish their performances based solely on force MAE.

In contrast, the MD stability trends diverge from the force MAE curve. Although the force MAE clearly decreases with further training, MD stability does not necessarily improve in tandem. In some cases, instability occurs earlier at later training epochs, reflected by a reduced instability onset time, despite better force prediction accuracy. This trend is particularly pronounced in the non-pre-trained case, suggesting that the model may overfit to the training dataset and struggle to generalize to diverse configurations encountered during extended MD simulations. This observation aligns with the findings of Fu et al.~\cite{fu2023forces}, indicating that force metrics alone may not reliably reflect downstream performance in MD simulations.

\subsection{Stability Enhancement}

Pre-training GemNet-T on the large and diverse OC20 dataset effectively mitigates overfitting and improves MD simulation stability. Although the reduction in force MAE is minimal, the impact on stability is far more pronounced. At epoch 300, the average instability onset time for the pre-trained model is 274.2\,ps with a standard deviation of 44.6\,ps, while the non-pre-trained model exhibits a significantly lower instability onset time of 42.6\,ps with a standard deviation of 25.9\,ps, as shown in Figure~\ref{fig:results}b. This stark contrast highlights the substantial improvement in long-term dynamical 
stability achieved through pre-training.
This indicates that the richer feature representations learned from OC20 effectively transfer to smaller tasks like MD17-Aspirin, enhancing the model’s ability to maintain stable configurations and avoid collapsing bond lengths or unphysical states, particularly during extended simulations.

\begin{figure}[!htb]
    \centering
    \includegraphics[width=0.99\textwidth]{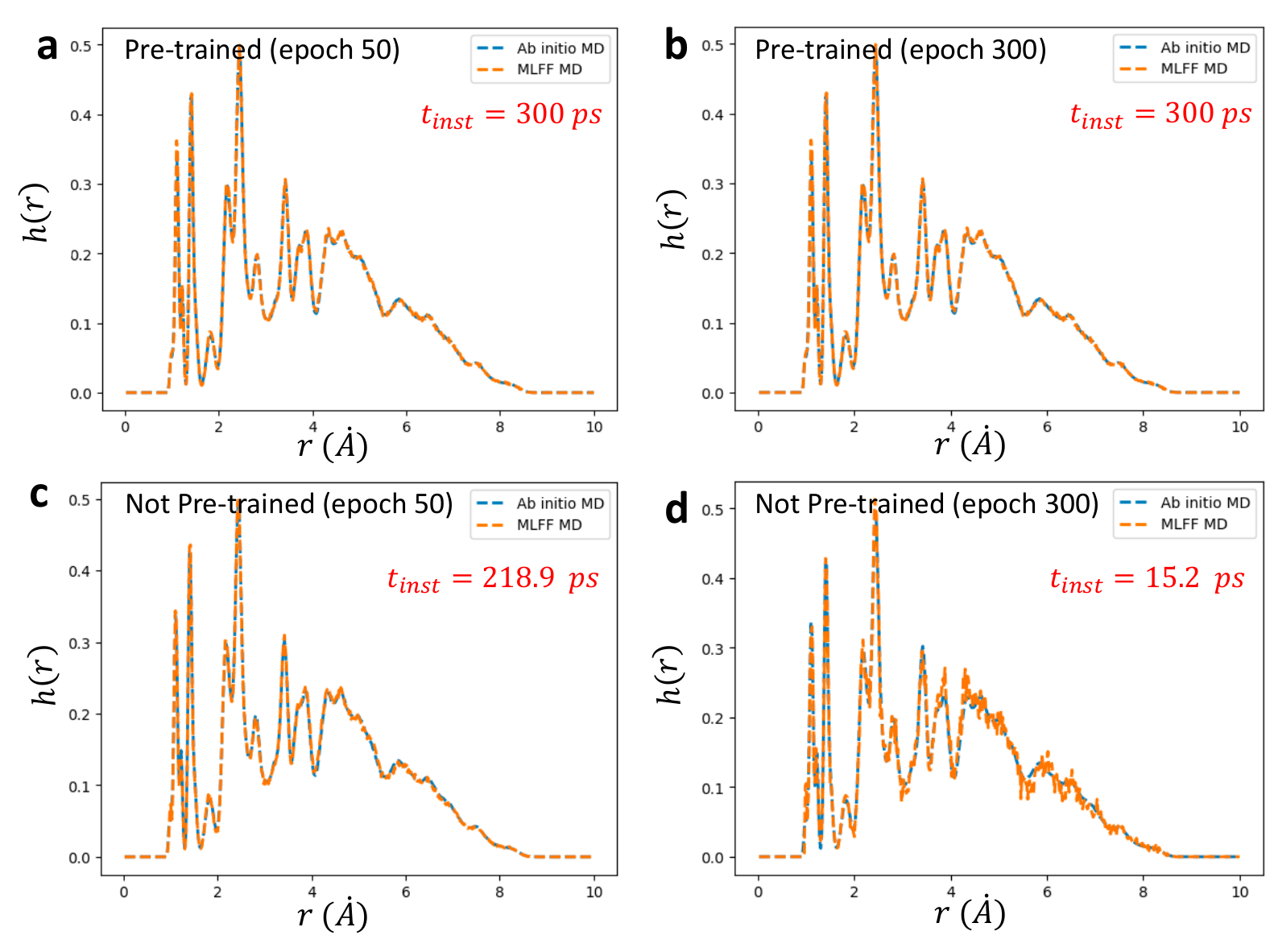}
    \caption{Pair-distance distribution comparisons for Aspirin. Each plot shows normalized interatomic distance distributions $h(r)$ from reference \textit{ab initio} MD (blue dashed) and MLFF-based MD (orange dashed) as a function of distance $r$ (\AA). Subplots \textbf{a.} and \textbf{b.} show results for pre-trained models at epochs 50 and 300; \textbf{c.} and \textbf{d.} show non-pre-trained models at the same epochs. Red labels indicate instability onset time (ps).}
    \label{fig:h(r)}
\end{figure}

A similar trend is observed in the pair-distance distribution function, as shown in Figure~\ref{fig:h(r)}. The model with pre-training maintains close agreement with the reference \emph{ab initio} MD simulation throughout training, whereas the non-pre-trained model exhibits growing deviations at later epochs. This divergence further highlights the stabilizing effect of pre-training on the structural fidelity of MD simulations.

\subsection{Better Latent Representations}

\begin{figure}[!htb]
    \centering
    \includegraphics[width=0.9\textwidth]{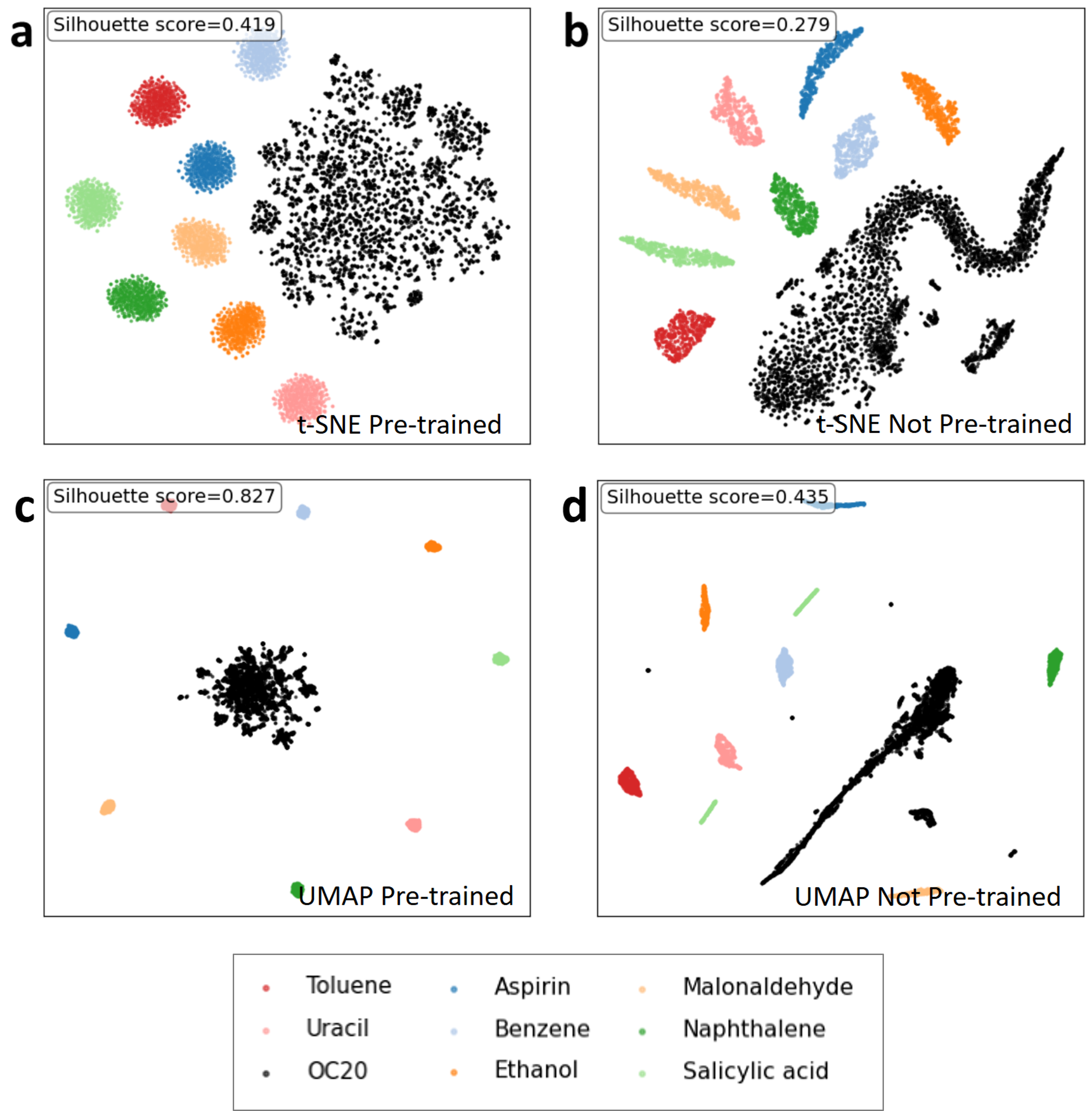}
    \caption{%
    2D projections of the learned embeddings for the eight MD17 molecular systems and OC20.  
    \textbf{a.} t-SNE projection for the pre-trained model.  
    \textbf{b.} t-SNE projection for the non-pre-trained model.  
    \textbf{c.} UMAP projection for the pre-trained model.  
    \textbf{d.} UMAP projection for the non-pre-trained model.  
    }
    \label{fig1}
\end{figure}

\begin{figure}[!htb]
    \centering
    \includegraphics[width=0.99\textwidth]{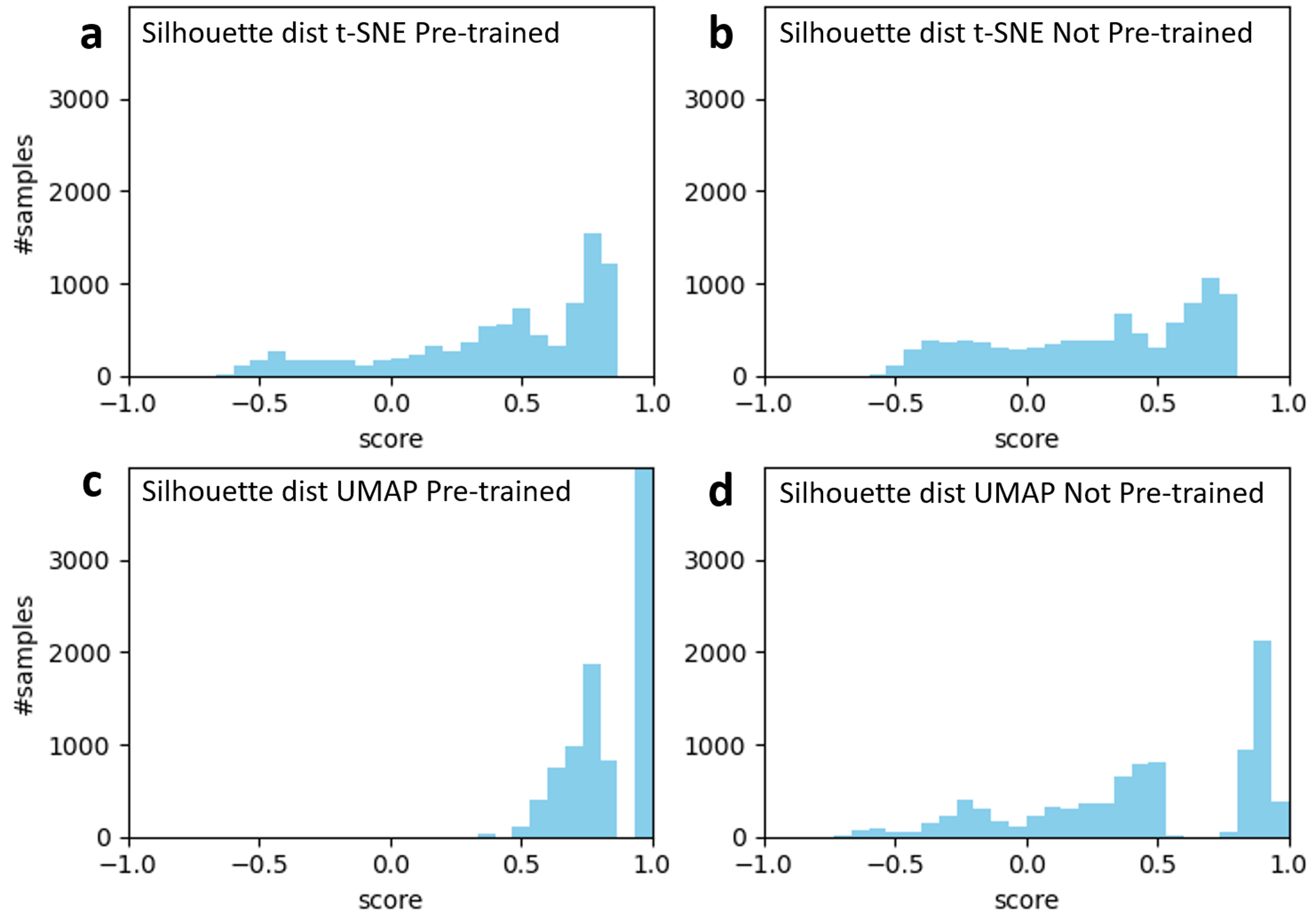}
    \caption{%
    Silhouette score statistics for the 2D projections in Figure~\ref{fig1}.  
    \textbf{a.} Distribution of per-sample silhouette scores for the t-SNE projection of the pre-trained model.  
    \textbf{b.} Distribution for the t-SNE projection of the non-pre-trained model.  
    \textbf{c.} Distribution for the UMAP projection of the pre-trained model.  
    \textbf{d.} Distribution for the UMAP projection of the non-pre-trained model.}
    \label{fig2}
\end{figure}

Since improved latent representations can enhance model performance, we investigated how pre-training affects the organization of atomic environment embeddings. Specifically, we concatenate the validation embeddings from the eight MD17 molecular systems and OC20 by stacking samples to form a joint embedding set, and project the resulting high-dimensional features into two dimensions using t-SNE and UMAP (Figure~\ref{fig1}a–d). t-SNE primarily focuses on preserving local neighborhoods and is particularly effective at revealing tight class-wise clusters~\cite{vandermaaten2008tsne}, whereas UMAP provides a stronger compromise between local neighborhood preservation and global manifold geometry, making the relative arrangement and shapes of clusters more interpretable~\cite{mcinnes2018umap}. Using both projections helps ensure that the qualitative conclusions are not tied to a single projection method but remain consistent under two widely used nonlinear embeddings. In the 2D embedding space, we compute the silhouette score as a measure of class separability and plot the distribution of per-sample silhouette values to compare the clustering quality of embeddings obtained with the two training strategies (Figure~\ref{fig2}a–d).

Under the t-SNE projection (Figure~\ref{fig1}a,b), the pre-trained model’s embeddings exhibit a clear class structure: each MD17 molecule forms a well-defined, approximately spherical cluster with compact within-class variation, while the OC20 configurations collapse into a single dense cloud in one region of the embedding space. This indicates that pre-training facilitates the learning of more discriminative representations. Samples from the same category become more coherent in feature space, while samples from different categories are more widely separated. This structure emerges even though the model was pre-trained only on OC20 catalyst systems and fine-tuned on the MD17 aspirin subset, without seeing the other MD17 molecules used in this analysis.

With UMAP (Figure~\ref{fig1}c,d), the pre-trained model likewise shows pronounced inter-class separability. Each MD17 molecule forms a compact, well-isolated island in the embedding space, while the OC20 configurations occupy one dense central region. Overall, the pre-trained model consistently displays a dense within-class and well-separated between-class organization under both projection methods, suggesting that the improved latent-space structure is robust rather than an artifact of a specific visualization technique.

In contrast, the non-pre-trained model exhibits visibly less sharp cluster boundaries in both t-SNE and UMAP. Several molecular classes appear as more irregular or band-like groups instead of tight blobs. Notably, in the UMAP projection the salicylic acid manifold is split into two separated streaks. The OC20 configurations also extend along an elongated main band and give rise to several small, isolated groups scattered across the embedding. These patterns suggest that the representations learned by the non-pre-trained model have weaker within-class cohesion and a less coherent geometric organization, even though large-scale overlaps between different molecular classes are still limited compared to the Pre-trained case.

To quantify class separability in the 2D projections, we use the silhouette score, which measures the relative separation between samples within the same class and those from different classes, and compute its mean over all samples in each setting (Figure~\ref{fig2}a–d). Higher values correspond to smaller average distances within the class and larger average distances between classes. The mean silhouette score provides a compact summary of overall clustering quality, while the full distribution of per-sample scores reveals how many points lie deep inside well-separated clusters versus near class boundaries or in ambiguous regions. In both projection methods, the pre-trained model achieves higher silhouette scores than the non-pre-trained model, indicating that pre-training substantially improves the discriminative structure of the representation space by yielding more compact within-class groupings and more clearly separated classes. We also examine the distribution of per-sample silhouette scores. The histograms (Figure~\ref{fig2}a–d) show that the pre-trained model has a higher count of samples in the high-score region (close to 1), whereas the non-pre-trained model exhibits a larger proportion of samples near zero or even negative values, reflecting a higher prevalence of boundary points and potentially misassigned samples. These distributional differences are consistent with the 2D scatter plots, where the pre-trained embeddings form more compact and largely single clusters, whereas the non-pre-trained embeddings tend to be more elongated and fragmented.

\subsection{Local Force Sensitivity via Bond Perturbation}

To assess how pre-training influences the physical regularity of the learned force field, we examine whether the model produces smooth and stable force responses under small, controlled geometric perturbations. In a physically meaningful potential energy surface, atomic forces are the (negative) gradients of the energy; therefore, when a single bond is compressed or stretched smoothly around equilibrium while all other coordinates are fixed, the corresponding projected interatomic force should vary smoothly with bond length. Moreover, the derivative \(dF_{\parallel}/dr\) reflects the local bond stiffness, and spurious high-frequency wiggles in this quantity typically indicate numerical artifacts or inconsistencies in the learned force field rather than true physical behavior.

To probe local force sensitivity, we perform a one-dimensional bond-length scan on selected covalent bonds in an aspirin molecule. For a target bond \((i,j)\), we cut the bond in a manually defined connectivity graph and identify the two resulting molecular fragments via a flood-fill search. We then rigidly translate one fragment (all atoms in the fragment are shifted by the same displacement) along the bond direction to systematically compress and stretch the bond length from \(0.95\,r_0\) to \(1.05\,r_0\) (21 geometries), while keeping the internal geometry of each fragment fixed. For each scanned geometry, we evaluate the model-predicted forces and analyze the bond-projected force response.

\begin{figure}[!htb]
    \centering
    \includegraphics[width=1.0\textwidth,trim=27mm 0mm 32mm 0mm, clip]{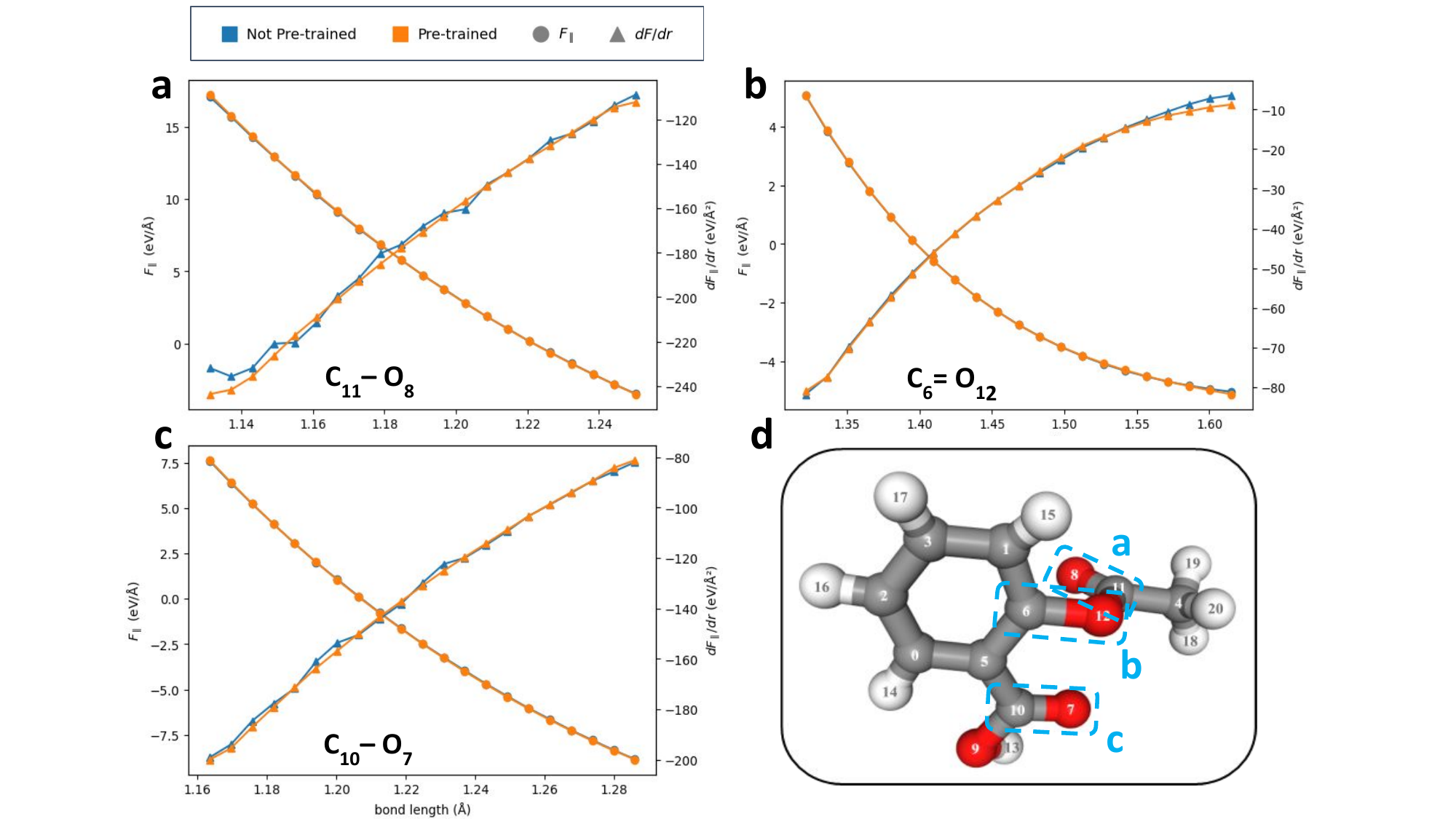}
\caption{
    Force--bond length profiles and their derivatives for three representative bonds in aspirin (atom indices follow the MD17 convention).
    Panels \textbf{(a--c)} correspond to: \textbf{(a)} a C--O single bond (C$_{11}$--O$_8$), \textbf{(b)} a C--O single bond (C$_6$--O$_{12}$), and \textbf{(c)} a C=O double bond (C$_{10}$=O$_7$).
    Panel \textbf{(d)} shows a schematic of the aspirin molecule highlighting the three selected bonds labelled \textbf{a}, \textbf{b} and \textbf{c}.
    For panels \textbf{(a--c)}, the projected bond force $F_{\parallel}(r)$ is shown versus bond length $r$ for the pretrained model and the model trained from scratch, together with the numerical first derivative      $dF_{\parallel}/dr$.
    }
    \label{fig3}
\end{figure}

Specifically, for a scanned bond between atoms \(i\) and \(j\), we define the bond vector \(\mathbf{r}_{ij} = \mathbf{r}_j - \mathbf{r}_i\), the bond length \(r = \lVert \mathbf{r}_{ij} \rVert\), and the unit vector \(\hat{\mathbf{u}}_{ij} = \mathbf{r}_{ij} / r\). Given the model-predicted atomic forces \(\mathbf{F}_i\) and \(\mathbf{F}_j\), the projected force along the bond is computed as
\begin{equation}
   F_{\parallel}(r) = \bigl( \mathbf{F}_j - \mathbf{F}_i \bigr)\cdot \hat{\mathbf{u}}_{ij}.
\end{equation}
We consider three representative bonds in aspirin: a C–O single bond (C\(_{11}\)–O\(_8\)), a C–O single bond (C\(_6\)–O\(_{12}\)), and a C=O double bond (C\(_{10}\)=O\(_7\)). For each bond, we scan \(r\) and numerically evaluate \(dF_{\parallel}/dr\) at each geometry. For all three bonds (Fig.~\ref{fig3}), the \(F_{\parallel}\)–\(r\) curves predicted by both models appear visually smooth, suggesting that both learn qualitatively reasonable force–bond length trends. However, clearer differences emerge in \(dF_{\parallel}/dr\): the pretrained model yields substantially smoother and more regular derivative curves, while the non-pretrained model exhibits pronounced small-scale fluctuations. This indicates that pre-training suppresses spurious oscillatory artifacts in the learned force field, leading to force responses that vary more smoothly with bond stretching and are therefore more physically consistent.

\subsection{Force Difference Prediction Enhancement}


MD advances a system through a sequence of closely related configurations, in which atoms undergo small displacements driven by thermal motion. The finite integration time step determines the magnitude of these step-to-step updates. In MLFF-based MD, small force prediction errors between adjacent frames can accumulate over time relative to reference MD, leading to trajectory drift, energy drift, or numerical instability even when the frame-wise force MAE appears acceptable \cite{chmiela2017machine,sauceda2019molecular}. To assess the propensity for such accumulation, we evaluate the mean absolute error of inter-frame force differences, \(\Delta\mathbf{F}\) MAE, for structurally similar configurations along the aspirin MD trajectory. Since discrete-time integration is driven by how forces change from one step to the next, \(\Delta\mathbf{F}\) MAE serves as a more direct proxy for the local dynamical consistency required for stable MLFF-based MD. Moreover, even when configuration-wise force errors are non-negligible, \(\Delta\mathbf{F}\) can remain accurate if the errors across adjacent frames are similarly directed or exhibit consistent trends, as these effects tend to cancel out in the difference \cite{ock2023beyond}.

We select a local subset of $N=100$ test configurations closest (by aligned RMSD) to a common reference configuration (frame id=21), forming a compact region of configuration space near the reference geometry. For each configuration, we summarize internal force variations using bond-projected force components along a fixed set of covalent bonds $\mathcal{B}=\{(p,q)\}$. Specifically, for bond $(p,q)$ in configuration $i$ with bond direction $\hat{\mathbf{u}}_{i;(p,q)}$, we define a scalar bond force
\begin{equation}
F^{\text{bond}}_{i;(p,q)} = \big(\mathbf{F}_{i}^{(q)}-\mathbf{F}_{i}^{(p)}\big)\cdot \hat{\mathbf{u}}_{i;(p,q)},
\end{equation}
and collect them into $\mathbf{F}^{\text{bond}}_{i}\in\mathbb{R}^{|\mathcal{B}|}$. For each unordered pair $(i,j)$, we compare the change in bond forces $\Delta\mathbf{F}^{\text{bond}}_{ij}=\mathbf{F}^{\text{bond}}_{j}-\mathbf{F}^{\text{bond}}_{i}$ predicted by the model to the corresponding DFT reference $\Delta\mathbf{F}^{\text{bond}}_{\text{ref};ij}$. The pairwise $\Delta\mathbf{F}$ MAE is
\begin{equation}
\mathrm{MAE}_{ij}
=\frac{1}{|\mathcal{B}|}\sum_{(p,q)\in\mathcal{B}}
\left|
\Delta F^{\text{bond}}_{ij,(p,q)}-
\Delta F^{\text{bond}}_{\text{ref};ij,(p,q)}
\right|.
\end{equation}
Averaging $\mathrm{MAE}_{ij}$ over all $i<j$ yields an overall local $\Delta\mathbf{F}$ error for this neighborhood.

\begin{figure}[!hptb]
    \centering
    \includegraphics[width=0.99\textwidth]{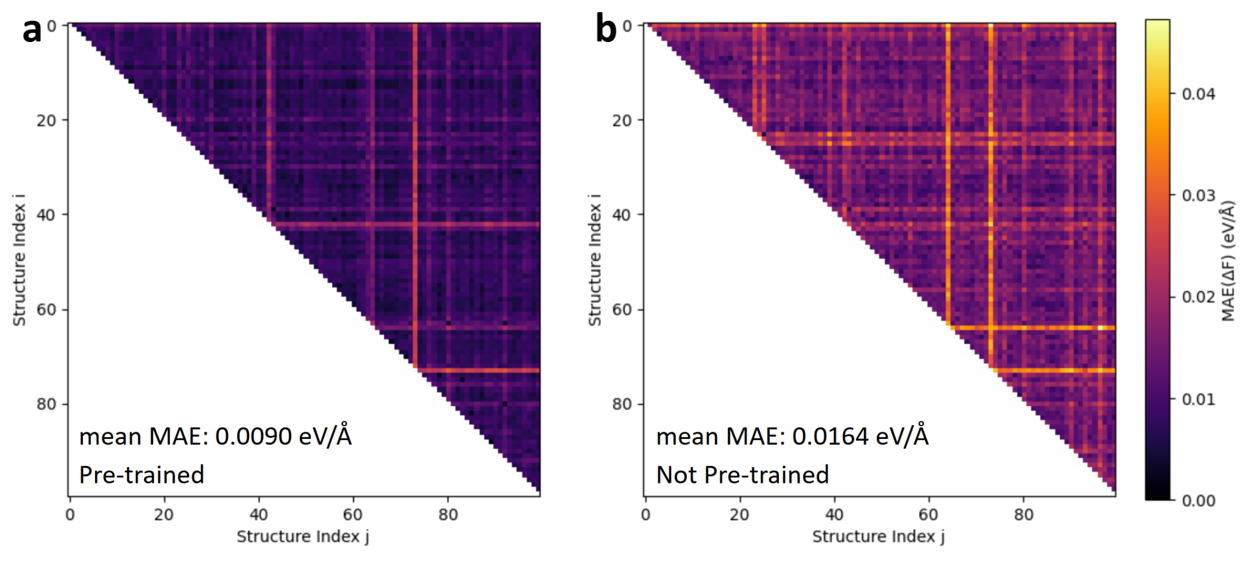}
    \caption{%
       Upper-triangular MAE matrices of pairwise bond-force differences for the pre-trained (a) and non–pre-trained (b) models on a subset of 100 similar aspirin structures.\\ Each axis denotes the structure indices \(i\) and \(j\) (\(0 \le i < j \le 99\)), and each pixel gives the \(\mathrm{MAE}_{ij}\) between the model-predicted and reference bond-force differences for the pair \((i,j)\). Darker colors indicate smaller errors and lighter colors indicate larger errors, with the pre-trained model exhibiting substantially smaller average errors over this region of configuration space.
    }
    \label{fig4}
\end{figure} 

The pre-trained model exhibits a uniformly darker matrix with fewer bright outliers (Figure~\ref{fig4}a), indicating consistently smaller errors in predicting relative bond-force changes across pairs of nearby configurations. In contrast, the model trained from scratch shows more pronounced light patches and sporadic high-error regions (Figure~\ref{fig4}b), suggesting less reliable local force variations even within this small RMSD neighborhood. Quantitatively, the mean $\Delta\mathbf{F}$ MAE within this neighborhood is 0.0090~eV/\AA{} for the pre-trained model, compared with 0.0164~eV/\AA{} for training from scratch. This reduction implies that pretraining improves the local consistency of the learned force field around the reference geometry, which is important for stable and physically reliable MD.

\section{Conclusion}

Building on the observation that force metrics alone may be insufficient for downstream applications of MLFF, such as MD simulations, we propose that pre-training MLFFs on large, diverse datasets can significantly enhance their performance in the MD simulation applications. Specifically, we compared a model trained solely on MD17 Aspirin data with one pre-trained on the broader OC20 dataset and then fine-tuned on the MD17 Aspirin subset. The pre-trained model demonstrates a distinct advantage in maintaining longer stable trajectories, particularly at later training epochs. We further investigate the mechanisms behind the improved stability brought by pre-training. Our analysis suggests that pre-training yields more structured latent representations, smoother force responses to local geometric perturbations, and more consistent force differences between nearby configurations. This improved robustness stems from the richer geometric and chemical diversity captured in the OC20 dataset, enabling the model to learn more transferable feature representations. This work lays the groundwork for developing more generalizable training strategies for MLFFs, ultimately advancing their applicability to diverse molecular systems and complex simulation environments.

\section*{Code Availability Statement}
The computational implementation and analysis code for this study are available in our public GitHub repository: \url{https://github.com/shagunm1210/MDsim-local}.

\bibliography{reference}

\end{document}